\documentclass{aa}  
\usepackage{graphicx}
\usepackage{txfonts}
\usepackage{lineno}
\begin{document}

   \title{Broad-band noises in GX 339-4 during the 2021 outburst observed with \textit{Insight-HXMT} and \textit{NICER}}

   \author{Yong-Jie Jin
          \inst{1,2}
          \and
          Xiao Chen\inst{1}
          \and
          Hai-Fan Zhu\inst{1}
          \and
          Ze-Jun Jiang\inst{2}
          \and
          Wei Wang\inst{1}\fnmsep\thanks{wangwei2017@whu.edu.cn}
          }

   \institute{Department of Astronomy, School of Physics and Technology, Wuhan University, Wuhan 430072, China
         \and
             Department of Astronomy, School of Physics and Astronomy, Yunnan University, Kunming 650500, China
             }

   \date{}

 
  \abstract
 { Rapid X-ray variability of GX 339$-$4 including the low-frequency quasi-periodic oscillations (LFQPOs) and broad-band noises have been observed with the Hard X-ray Modulation Telescope (\textit{Insight}-HXMT) and Neutron star Interior Composition Explorer (\textit {NICER}) during the 2021 outburst. Here we present a systematic study of the evolution and energy dependence properties of such broad-band noises (BBN). The outburst from February to March of 2021 can be divided into three stages: the low hard state (LHS), the hard intermediate state (HIMS) and soft intermediate state (SIMS). In the PDSs of the LHS and HIMS, the broad-band noises are well fitted with three Lorentzian components: a low-frequency component $L_1$, a middle-frequency component $L_2$ and a high-frequency component $L_3$. The increasing trend of the characteristic frequencies for $L_1$ and $L_2$ and the relation between the QPO frequency and characteristic BBN frequency are reported. We found that the energies corresponding to the peaks and shapes of the rms spectra for three BBN components are different. The comparison among three BBN components indicates that energy-dominant bands of these BBN components are distinct. Our results can be explained with the truncated disc/hot flow model with a large variable disc and a small hot inner flow. A possible description of the accretion structure and its evolution from the LHS to the SIMS is proposed. Further research is still required to probe such accretion structure in GX 339--4.}

   \keywords{accretion: accretion discs -- black hole physics -- X-rays: binaries
               }

   \maketitle
%

\section{Introduction}
\label{sec:in}
During X-ray outbursts of black hole X-ray binaries (BHXRBs), vast amounts of energy are released making them the brightest sources in the X-ray sky. At the beginning of the outburst, a BHXRB is typically observed in the low/hard state (LHS) with increasing luminosities, where the energy spectrum is dominated by a hard power-law component (photon index $< 2$). As the luminosities reach the peak, the outburst enters the intermediate state (IMS) with a soft thermal component and a hard non-thermal component both contributing significantly to the energy spectrum. The IMS is also divided into the Hard Intermediate State (HIMS), which appears after the LHS, and the Soft Intermediate State (SIMS) with the softer energy spectrum than the HIMS. With the increasing contribution from the thermal component of the accretion disc, the outburst finally reaches the high/soft state (HSS). After that, the outburst evolves back to quiescence through the IMS again. A counter-clockwise Q-shaped loop in the hardness-intensity diagram (HID) is used to track the complete evolution of the X-ray outburst of BHXRBs \citep[see][for reviews]{Belloni2005,Remillard2006,Belloni2016}

The rapid X-ray variability in the outburst exists in the light curves and is generally studied in the Fourier domain with the power density spectrum (PDS). The low frequency quasi-periodic oscillation (LFQPO) is one of the important features of the rapid variability in PDSs, and can be classified into Types A, B, C \citep{Remillard2002,Motta2011,Motta2015,Ingarm2019}. Type-C QPOs are the most common type of QPOs in BHXRBs, and characterized by a strong, narrow peak superposed on a broad-band noise that steepens above a frequency comparable to the QPO frequency \citep{Casella2005}. The broad-band noise is characterized by a wide-frequency continuum component with a low-frequency and a high-frequency break. The properties of the LFQPOs and BBN show a correlation with the outburst state of the source \citep{Done2007}. As the source softens from the LHS to the HIMS, the QPO frequency and the low-frequency break of the broad-band noise increase to higher frequency \citep{Klis2004} and a correlation between them was found by \cite{Wijnands1999}.

Most of the spectral evolution mentioned above can be well explained in the truncated disc/hot flow geometry \citep{Schnittman2006,Ingram2009,Ingram2011}, which assumes a hot, geometrically thick, optically thin inner accretion flow and a cool, geometrically thin, optically thick, outer accretion disc truncated at a large radius. As the truncated radius of the accretion disc decreasing in the transition from the Compton dominated state (LHS) to the disc dominated state (HSS), the characteristic time-scale of all rapid variability shows a decreasing trend, which is most obvious in the evolution of the low-frequency break of the broad-band noise and QPO frequency \citep{Wijnands1999}. The broad-band noises can be explained with fluctuations in mass accretion rate produced in the viscous time scale of the outer region of flow propagating inwards through the entire hot flow to reach the innermost region close to the BH \citep{Uttley2001,Uttley2004,Uttley2005}, while the QPO could originate from the Lense-Thirring (L-T) precession of the entire hot flow \citep{Schnittman2006,Ingram2009,Ingram2010,Ingram2015,Motta2015}. It is worth noting that the L-T precession model is not the only explanation for the origin of LFQPOs \citep{Marcel2021}. Many other physical models, including the accretion ejection instability (AEI) model \citep{Tagger1999,Varnière2002},the propagating oscillatory shock (POS) model \citep{Molteni1996,Chakrabarti2008,Chakrabarti2015,Chatterjee2016} and the dynamic corona-disc coupling model \citep{Mastichiadis2022}, have been proposed to explore the possible origin of LFQPOs.

In the fluctuation propagation model\citep{Lyubarskii1997,Ingram2012,Ingram2016,Mushtukov2019}, perturbation occurs at each radius of the accretion flow. Slow fluctuations are produced at the large radius and propagate inwards. With the fluctuation propagating inwards, the PDS at any radius carries not only the feature of the fluctuations generated in the corresponding radius, but also the imprint of variability generated at all larger radii. Therefore, even the slowest fluctuation from the outer region will modulate variability of the inner region with propagating through the entire flow. The low-frequency break of the broad-band variability (the slowest fluctuations) is corresponding to the viscous time scale at the outer radius of hot flow. However, the high-frequency break is not simply corresponding to the viscous time scale at the inner edge of the hot flow. A critical time scale capable of generating coherent fluctuations needs be taken into count, because fluctuations on time scales shorter than the propagation time are incoherent \citep{Ingram2011}.

\cite{Kawamura2022} proposed a spectral-timing model including both the propagating fluctuations and reverberation based on the truncated disc/hot flow geometry. The variable flow included one thermal component of the variable disc and two Comptonisation components of the hot flow. This hot flow is inhomogeneous and radially stratified, and has a discontinuous jump with the variable disc in viscous time scale. The double-hump shape in PDSs and the increasing high-frequency variability with energy could be naturally explained with this model. \cite{Yang2022} studied the long-term evolution of the broad-band noise variability with more than 40 days data for MAXI J1820+070 and investigated the energy dependence of the broad-band noise with broad energy range from 1-150 keV. Their results also suggested the inhomogeneous hot flow and the geometry with a truncated accretion disc and two Comptonization regions. \cite{Kawamura2023} also applied the spectral-timing model on the PDSs of the HIMS where only one single high frequency hump rather than double-hump was found, and proposed the variable disc merged with the outer region of hot flow when it was close to the ISCO. Combined with the geometry proposed for the LHS in \cite{Kawamura2022}, they gave a smooth transition of the accretion geometry from the LHS to the HIMS.

GX 339$-$4 is a low-mass black hole X-ray binary discovered in 1973 \citep{Markert1973}, with a mass function of f(M) = 5.8 $\pm 0.5 M_{\sun}$ \citep{Hynes2003}, the distance of 6-15 kpc \citep{Hynes2004} and the relatively low inclinations ($i \leq 45^\circ$, \citealt{Nowak2002,Hynes2003}). GX 339$-$4 has undergone a number of outbursts in the past thirty years and been studied at all wavelengths. A new outburst of GX 339$-$4 in 2021 was observed by \textit{Insight}-HXMT. A complete HID was shown by \cite{Liu2023}, which suggested this outburst as a complete outburst with all accretion states.

During the LHS and the HIMS, type-C QPOs with its harmonic component and broad-band noises were found in the PDSs \citep{Jin2023,Liu2023,Aneesha2024}. The energy dependence of two prominent broad-band components was studied with AstroSat observations\citep{chand2024,Hitesh2024}. \cite{Stiele2023} made a detailed study of the evolution of the PDSs from the HIMS to the SIMS, in which the appearance and disappearance of Type-C and Type-B QPOs were observed. The different properties of Type-C and Type-B QPOs and the associated change in spectral properties suggest that they may originate from two distinct physical mechanisms. Type-B QPOs with its sub-harmonic and harmonic components were more prominent in the low energy bands \citep{Mondal2023}. The energy-dependent results of rms spectra and time lag suggest that Type-B QPOs may be associated with the corona or jet. \cite{Valentina2023} found that the dual-corona model could give a better fitting for the rms and phase lag spectrum of the Type-B QPOs in GX 339-4, and believed that the corona could interact with the accretion disc by the feedback of X-ray photons. \cite{Yang2023} made a broad-band spectral-timing analysis for the periods with broad-band noise and Type-B QPOs, and suggested that a truncated disc/hot flow geometry exists during the period with broad-band noise. The hot flow region may convert into a precessing jet with Type-B QPOs appearing, whereas the jet stops precessing leading to the disappearance of Type-B QPOs.

In this paper, we study the evolution of broad-band noises in GX 339$-$4 from the low hard state to the soft intermediate state of the 2021 outburst based on Insight-HXMT and NICER observations. The relation between QPOs and broad-band noises and the energy dependence of broad-band noises are also studied. In Sect.~\ref{sec:ob}, we describe the observations and data reduction methods. Our main results are presented in Sect.~\ref{sec:re}, and discussions on the implications and model are shown in Sect.~\ref{sec:dis}. The conclusions follow in Sect.~\ref{sec:co}.

\section{Observation and data analysis}
\label{sec:ob}

\textit {Insight}-HXMT is China's first X-ray satellite, which is capable of spectral and timing observations from 1 keV to 250 keV \citep{Zhang2020}. It carries three detectors: the High-Energy X-ray telescope (HE), the Medium-Energy X-ray telescope (ME), and the Low Energy X-ray telescope (LE), covering the energy band of 20-250 keV, 5-30 keV and 1-15 keV, respectively. HE detector is made by 18 NaI/CsI detectors and has large area of $5100 \rm cm^2$. ME detector has 1728 Si-PIN detectors with $952\rm cm^2$ area, while LE detector is a Swept Charge Device with areas of $384\rm cm^2$. The Neutron star Interior Composition Explorer (\textit {NICER};\citealt{Gendreau2016}), which is an X-ray telescope on the International Space Station (ISS), is sensitive for the energy band of 0.2-12 keV and has the effective area of $> 2000\rm cm^2$ at 1.5 keV and $600 \rm cm^2$ at 6 keV. It has good energy resolution, timing resolution, and sensitivity in the soft X-ray bands.

\begin{figure}
    \centering
        \includegraphics[width=\columnwidth]{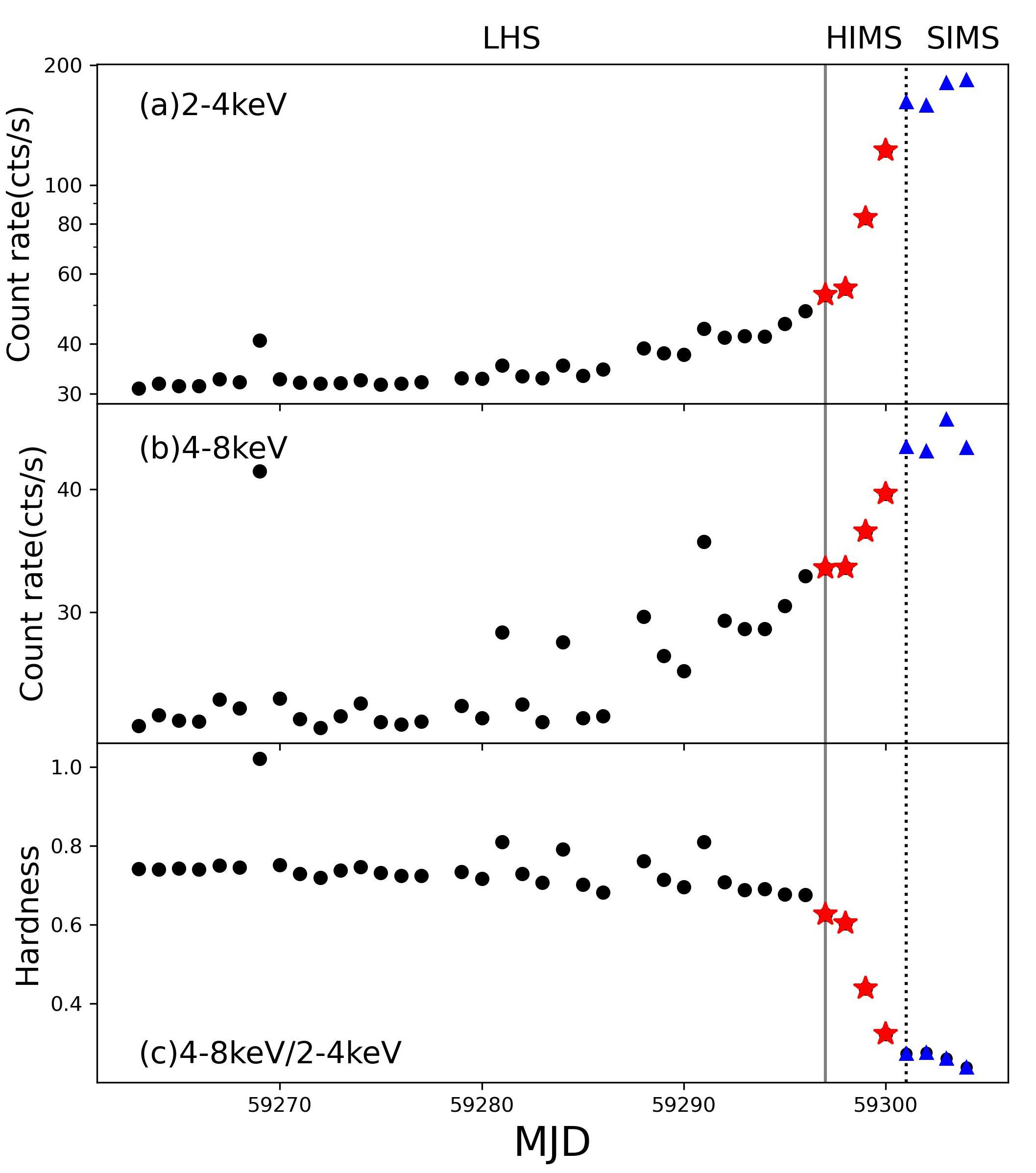}
    \caption{The count rate curves of GX 339-4 with  \textit {Insight}-HXMT LE ($2-4$ keV) and ($4-8$ keV) data are shown in the panel (a) and (b), respectively. Each point represents one day. The red stars represent the days in the HIMS and the blue triangles represent those in the SIMS. The hardness ratios are calculated with ($4-8$ keV) / ($2-4$ keV) and shown in the panel (c). The solid line marks the source in the transition from the LHS to the HIMS, and the dotted line marks the source in the transition to the SIMS.}
    \label{fig:light curve}
\end{figure}

GX 339$-$4 entered a new outburst in 2021 and was observed by \textit{Insight}-HXMT from February 18. In this work, we analysed the \textit{Insight}-HXMT data from February to March with Insight-HXMT Data Analysis software (HXMTDAS) v2.04 (see count rate curves and hardness ratio in Fig. ~\ref{fig:light curve}). The \textit{Insight}-HXMT data were reduced with the following criteria: the pointing offset angle less than $0.04^\circ$ , the elevation angle larger than $10^\circ$ and geomagnetic cutoff rigidity larger than $8^\circ$ . Data within 300 s of the South Atlantic Anomaly (SAA) passage were not used. The light curves were made by HXMTDAS tasks \textit{helcgen}, \textit{melcgen} and \textit{lelcgen} with 0.0078125s time bin. Backgrounds of both spectra and light curves were estimated with the official tools (\textit{HEBKGMAP}, \textit{MEBKGMAP} and \textit{LEBKGMAP}) of version 2.0.12. 

NICER also has lots of the observations for the 2021 outburst of GX 339-4, but only the simultaneous NICER observations of the \textit{Insight}-HXMT observations were analysed in this work. The NICERDAS ver.10 software was used with the calibration files, CALDB xti20221001. The light curves were generated by the NICER standard pipeline with the tasks nicerl2 and nicerl3-lc. The time bins of NICER light curves are 0.0078125s.

To obtain a longer exposure time, we used light curves of all observations within one day rather than each observation to produce the PDSs. The data intervals are set to 128 s while the time resolution is 1/128 s, corresponding to a Nyquist frequency of 64 Hz. The PDS was subjected to Poisson noise subtraction and fractional rms normalization \citep{Miyamoto1991}. We fit the PDSs with multiple Lorentzian functions using XSPEC v12.12.1. One of these Lorentzian components corresponds to the QPO frequency, and broad-band noises can usually be well described with 2-4 Lorentzian components \citep{Dziełak2021,Kawamura2022}. In this work, we found the broad-band noise of GX 339-4 can be well fitted with three Lorentzian components: a low-frequency component $L_1$, a middle-frequency component $L_2$ and a high-frequency component $L_3$ ( see Fig.~\ref{fig:qpo_com}). The characteristic frequency $v_{max}$, defined as $\sqrt{v_0^2+(FWHM/2)^2}$, where $v_0$ is the centroid frequency and FWHM is the full width at half maximum of the Lorentzian function, and the fractional rms amplitude were calculated to describe each noise component. The centroid frequency and the fractional rms amplitude of QPOs were also calculated so that the correlation between broad-band noises and QPOs can be studied.
The correlation between broad-band noises and QPOs is determined by linear regression. The slope of the linear regression curves and the Pearson correlation coefficients (PCCs) are calculated to determine the significance levels of linear correlations. 

To study the energy dependence of three components, the PDSs were produced in X-ray light curves of eight energy bands determined together with NICER (0.5-1 keV, 1-1.5 keV, 1.5-2 keV, 2-4 keV, 4-8 keV) and \textit{Insight}-HXMT (10-30 keV, 28-50 keV, 50-100 keV ) data. We found that the prominent QPOs were only observed in some energy bands after energy segmentation. The appearance and disappearance of the QPOs would affect the constraint of three broad-band components in different energy bands. Hence, we selected one representative day without prominent QPOs observed for the energy dependence study of broad-band noises based on the following criteria. For both NICER and \textit{Insight}-HXMT data, (a) the total exposure time within this day is more than 4000 s; (b) three broad-band components are prominent in all energy bands and (c) no prominent QPO is observed.

\section{Results}
\label{sec:re}

\begin{figure}
    \centering
        \includegraphics[width=\columnwidth]{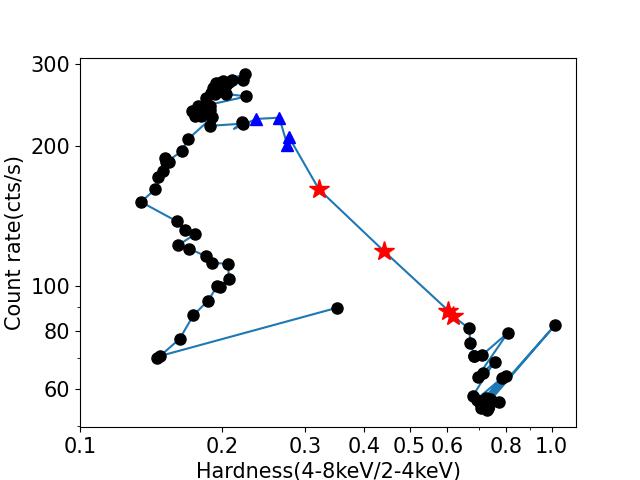}
    \caption{The hardness intensity diagram of GX 339$-$4 during the 2021 outburst. The Y-axis represents the count rate of LE from $2-8$ keV. The hardness ratio (X-axis) is defined as the count ratio between the energy bands $2-4$ keV and $4-8$ keV. Each point represents one day. The red stars represent the days in the HIMS and the blue triangles represent those in the SIMS.}
    \label{fig:HID}
\end{figure}

\begin{figure}
    \centering
        \includegraphics[width=\columnwidth]{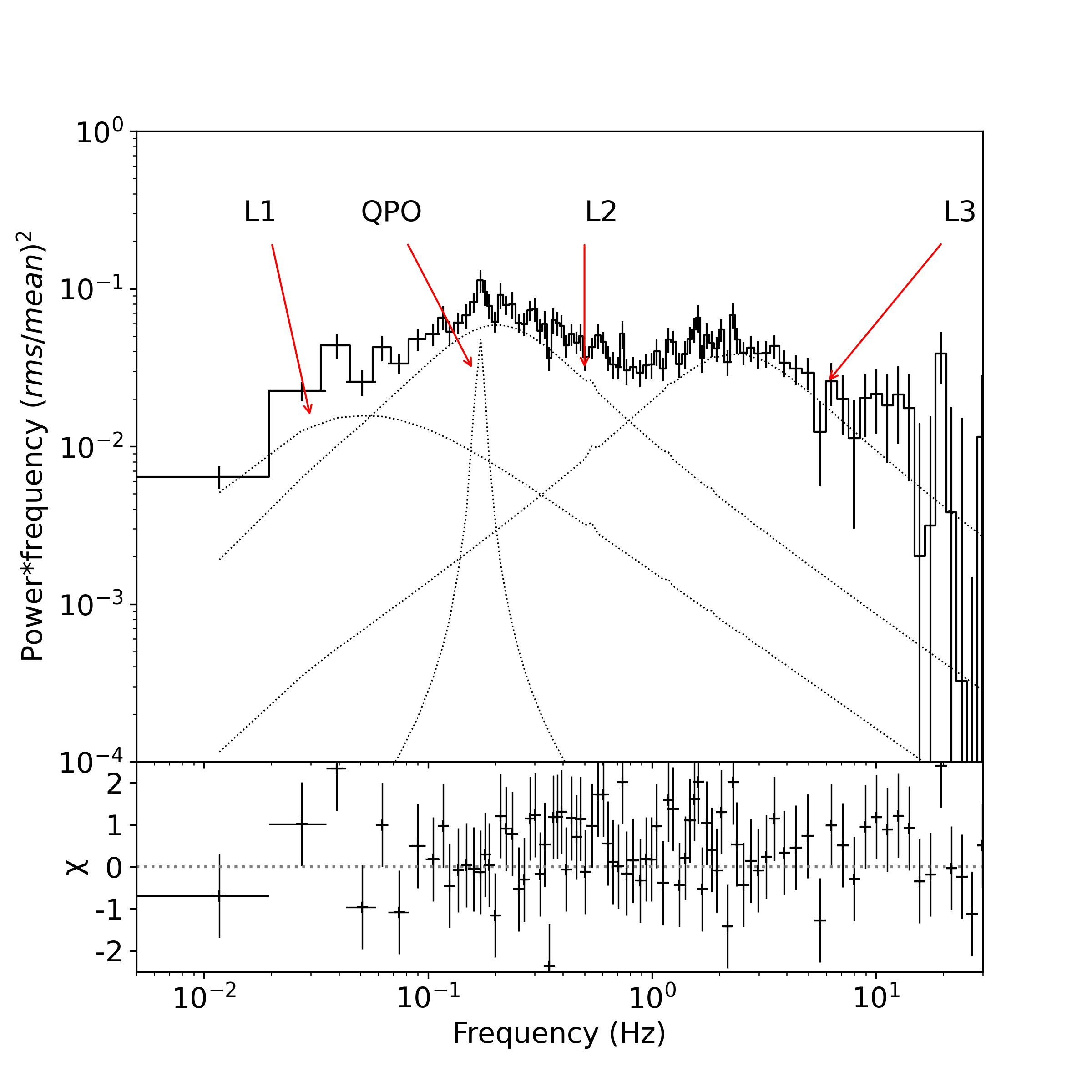}
    \caption{The representative power density spectrum with the Insight-HXMT/LE (1-10 keV) data. The dotted lines show the best fit with each component of a multiple-Lorentzian function. The QPO component has been shown while the three broad-band noise components are represented with the $L_1$, $L_2$ and $L_3$. }
    \label{fig:qpo_com}
\end{figure}

\begin{figure*}
    \centering
        \includegraphics[width=\linewidth]{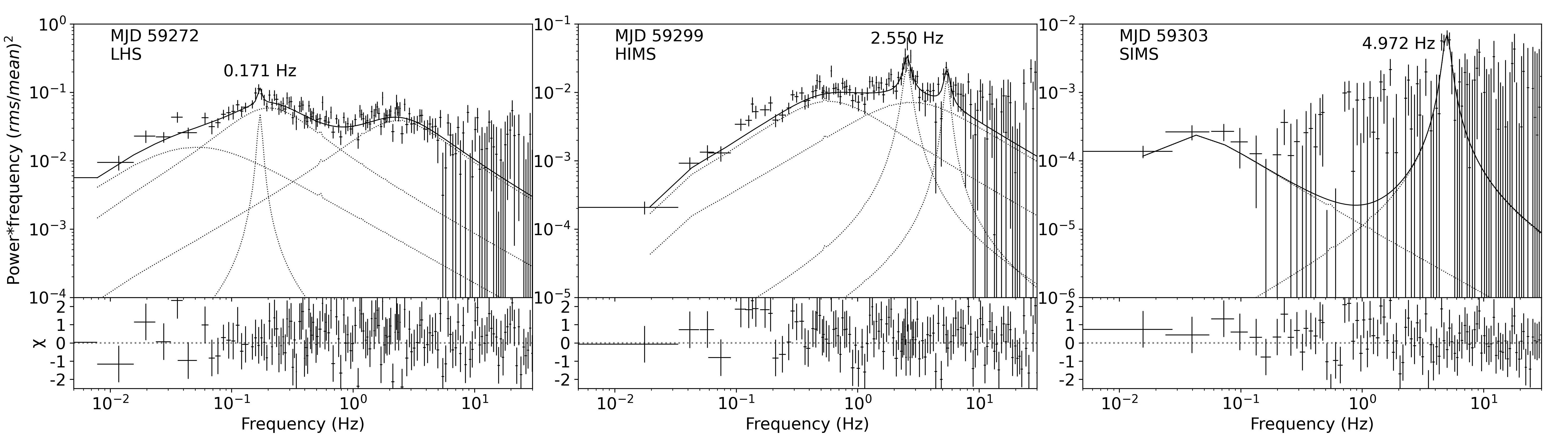}
    \caption{ The representative power density spectra for the different outburst stages using the Insight-HXMT/LE data (1-10 keV). The solid lines show the best fit with a multiple-Lorentzian function (dotted lines). The MJD, the corresponding outburst stage and the QPO fundamental frequency are shown for each panel. The left, middle and right panel show the PDS in the LHS, the HIMS and the SIMS, respectively.}
    \label{fig:qpo_stage}
\end{figure*}

\begin{figure*}
    \centering
        \includegraphics[width=\linewidth]{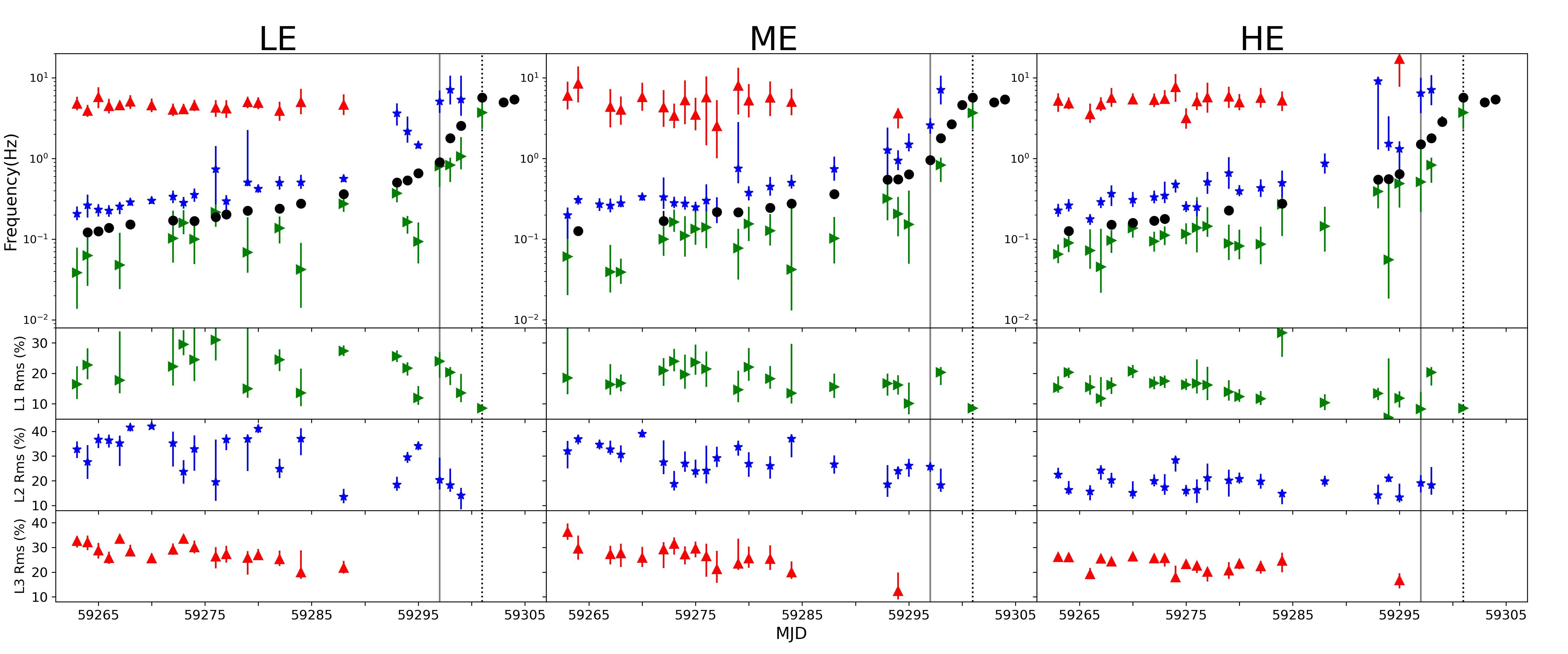}
    \caption{The evolution of the frequency and fractional rms of four components with time for Insight-HXMT/LE (1-10 keV, left column), ME (10-30 keV, middle column) and HE (28-100 keV, right column) data. In top row of panels, the evolution of the frequency is shown with black points for the QPO component and green right-triangles, blue stars and red up-triangles for the $L_1$, $L_2$ and $L_3$ components, respectively. In the three rows of panels below, the evolution of fractional rms of the $L_1$, $L_2$ and $L_3$ components with time has shown.}
    \label{fig:time}
\end{figure*}

\begin{figure*}
    \centering
        \includegraphics[width=0.85\linewidth]{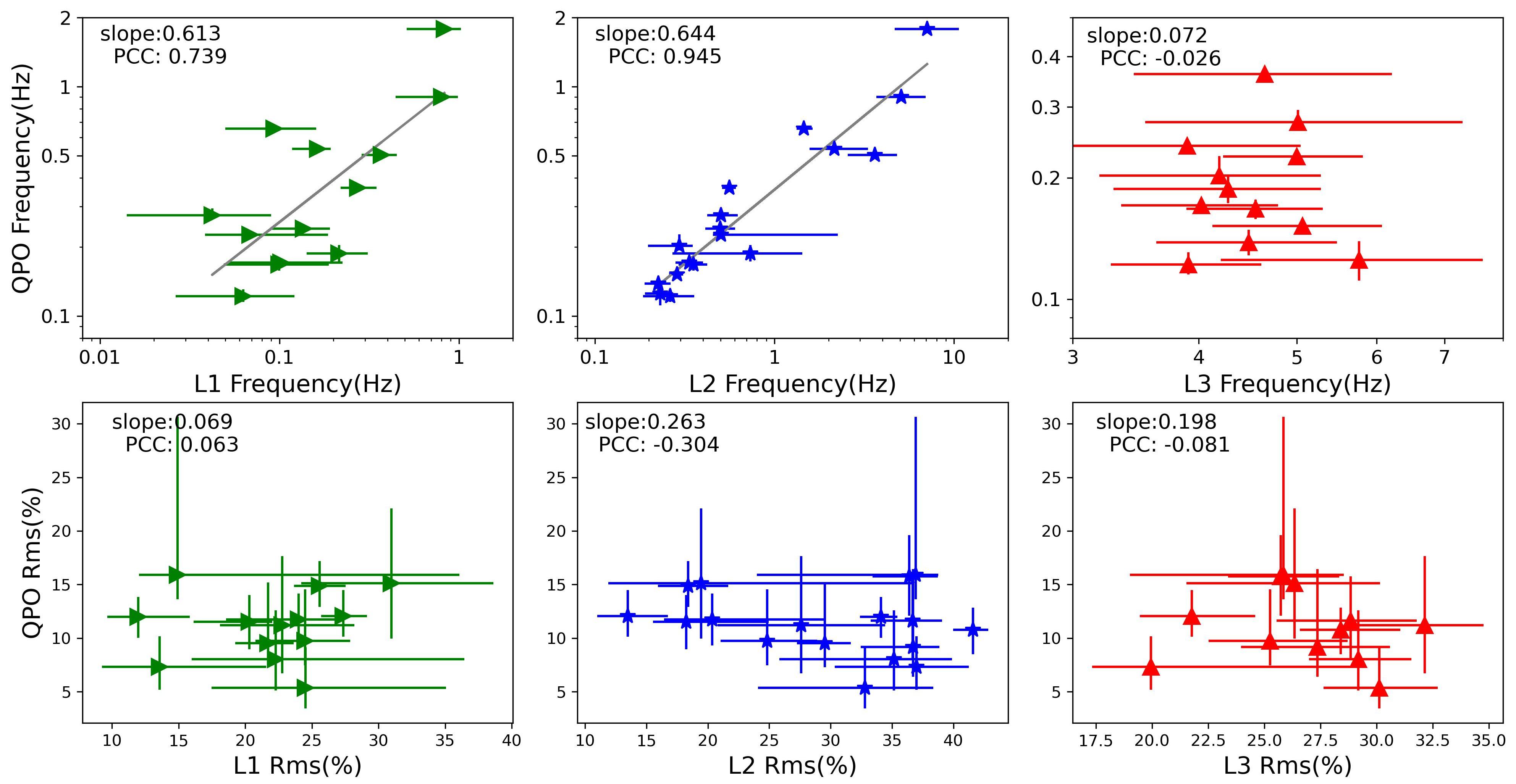}
    \caption{The relation between the QPO components and the $L_1$, $L_2$ and $L_3$ components. The frequency relation has shown in the top row of panels, while the rms relation has shown in the bottom row of panels. The green right-triangles, blue stars and red up-triangles represents the $L_1$, $L_2$ and $L_3$ components. The slope of linear regression curve and Pearson correlation coefficients (PCCs) have been shown in the panels. The solid lines are the linear regression curves for QPO frequencies versus L1 and L2 frequencies in the logarithmic axis. }
    \label{fig:relation}
\end{figure*}

\begin{figure*}
   \centering
        \includegraphics[width=1.35\columnwidth]{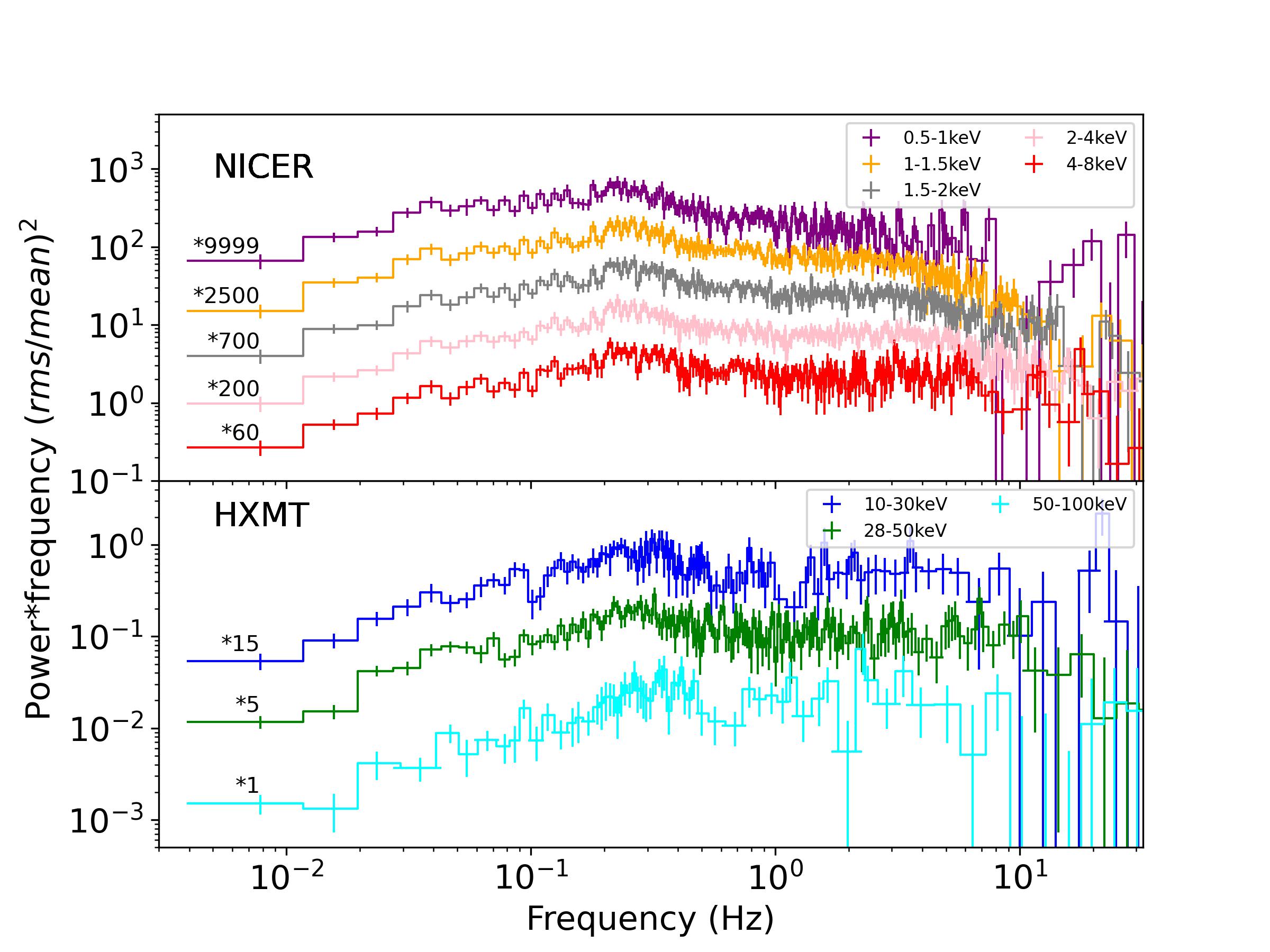}
    \caption{The power density spectra with the NICER (top panel) and Insight-HXMT (bottom panel) data in MJD 59280. From the top to the bottom is the PDS with 0.5-1 keV, 1-1.5 keV, 1.5-2 keV, 2-4keV, 4-8 keV, 10-30 keV, 28-100 keV data, respectively. The PDSs have been shifted vertically with different factors to show the comparison between the different photon energies.}
    \label{fig:qpo_energy}
\end{figure*}

\begin{figure*}
    \centering
        \includegraphics[width=0.85\linewidth]{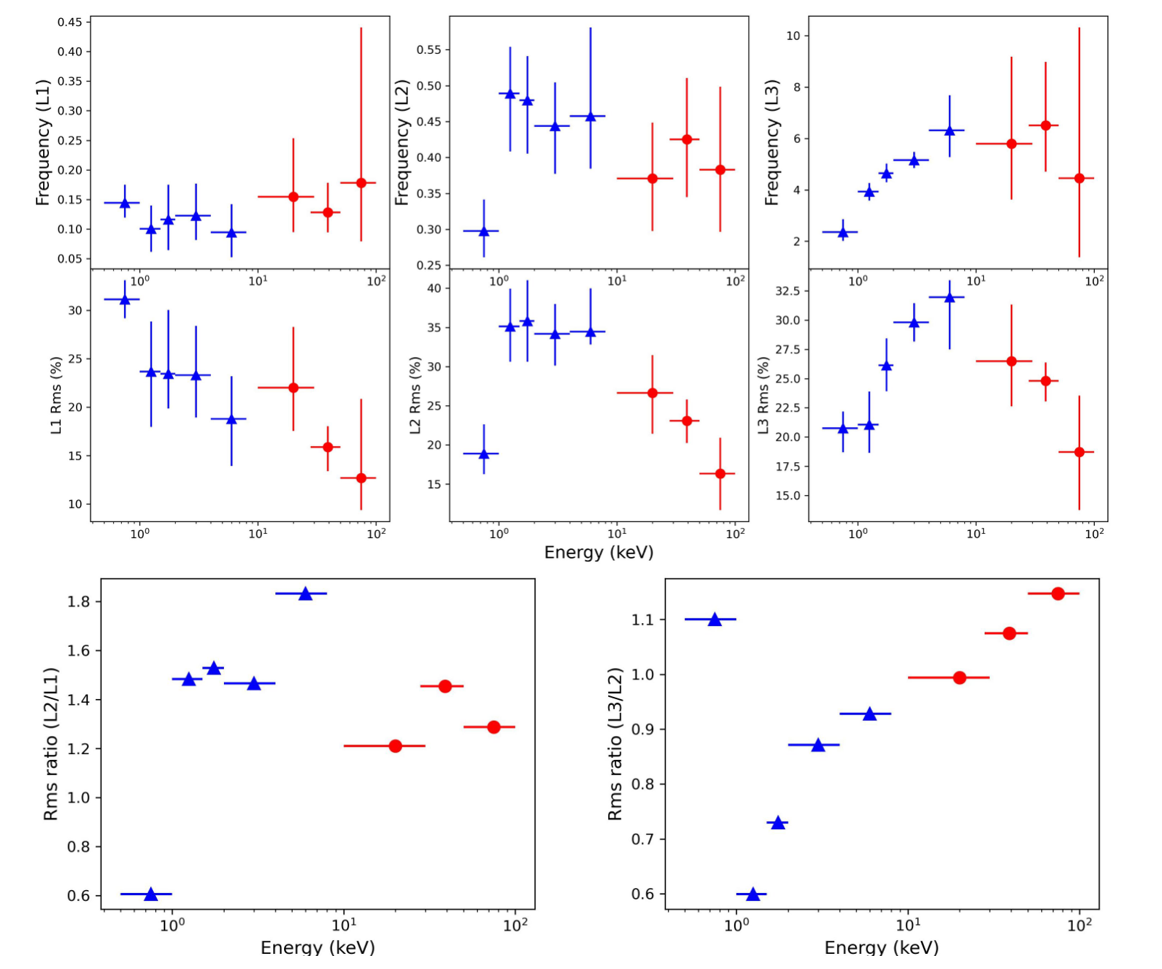}
    \caption{  The evolution of the frequency (top row), fractional rms (middle row) of the $L_1$,$L_2$ and $L_3$ components, and the ratio of the fractional rms (bottom row) between $L_1$ and $L_2$, $L_3$ and $L_2$ with photon energy. The blue triangles and red points represent the NICER and Insight-HXMT data, respectively. }
    \label{fig:energy}
\end{figure*}

\begin{figure}
        \includegraphics[width=\columnwidth]{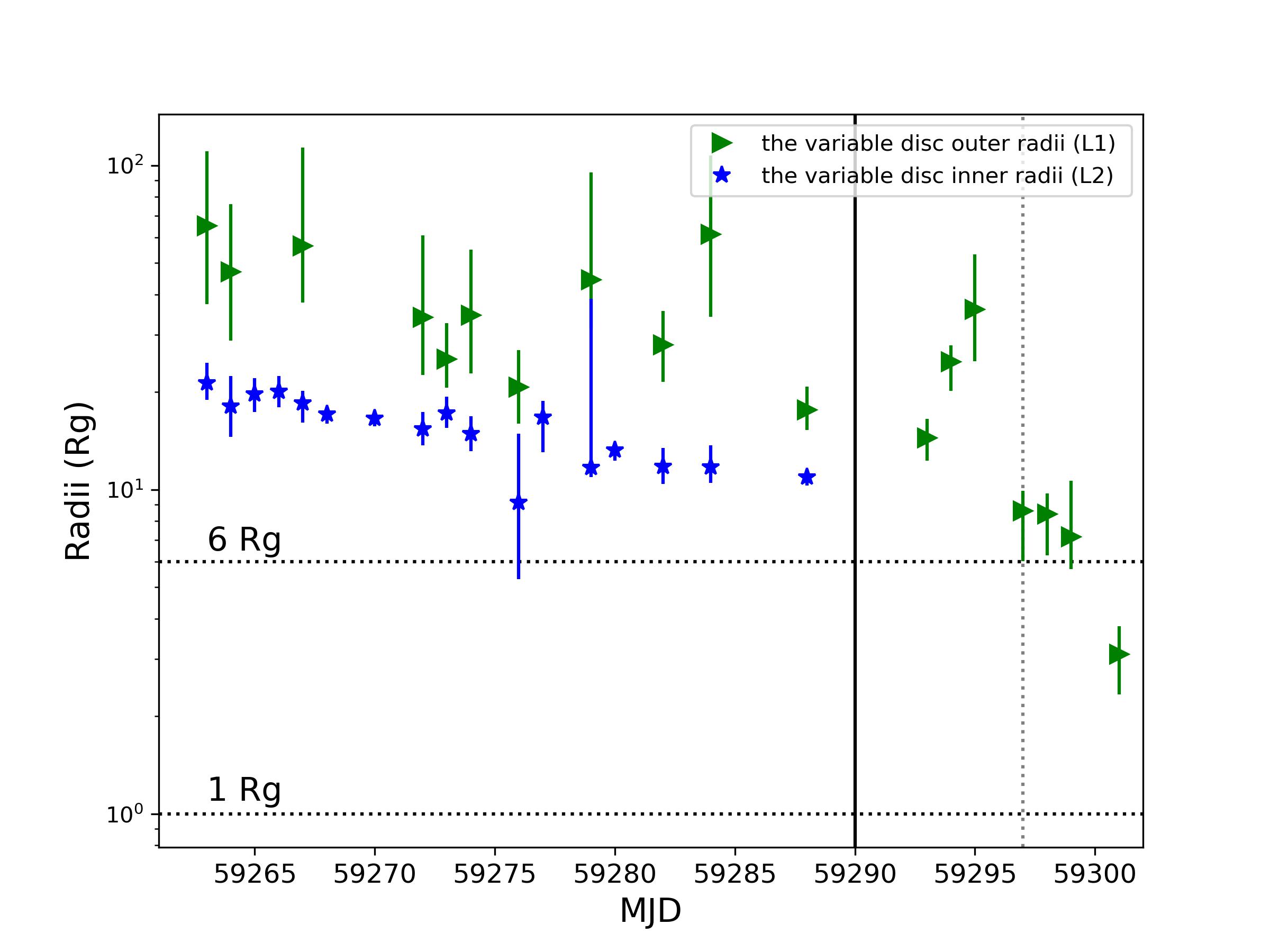}
    \caption{The evolution of the inner (green right-triangles) and outer (blue stars) radii of the variable disc with time. The black vertical solid line represents when the L3 component disappeared and the vertical dotted line represents when the outburst entered the SIMS. The ISCO positions for the black hole (a=0, $6R_g$) and fast-spinning black hole (a=1, $1R_g$) have been shown with the horizontal dotted lines.}
    \label{fig:radii}
\end{figure}

\begin{figure*}
   \centering
        \includegraphics[width=1.25\columnwidth]{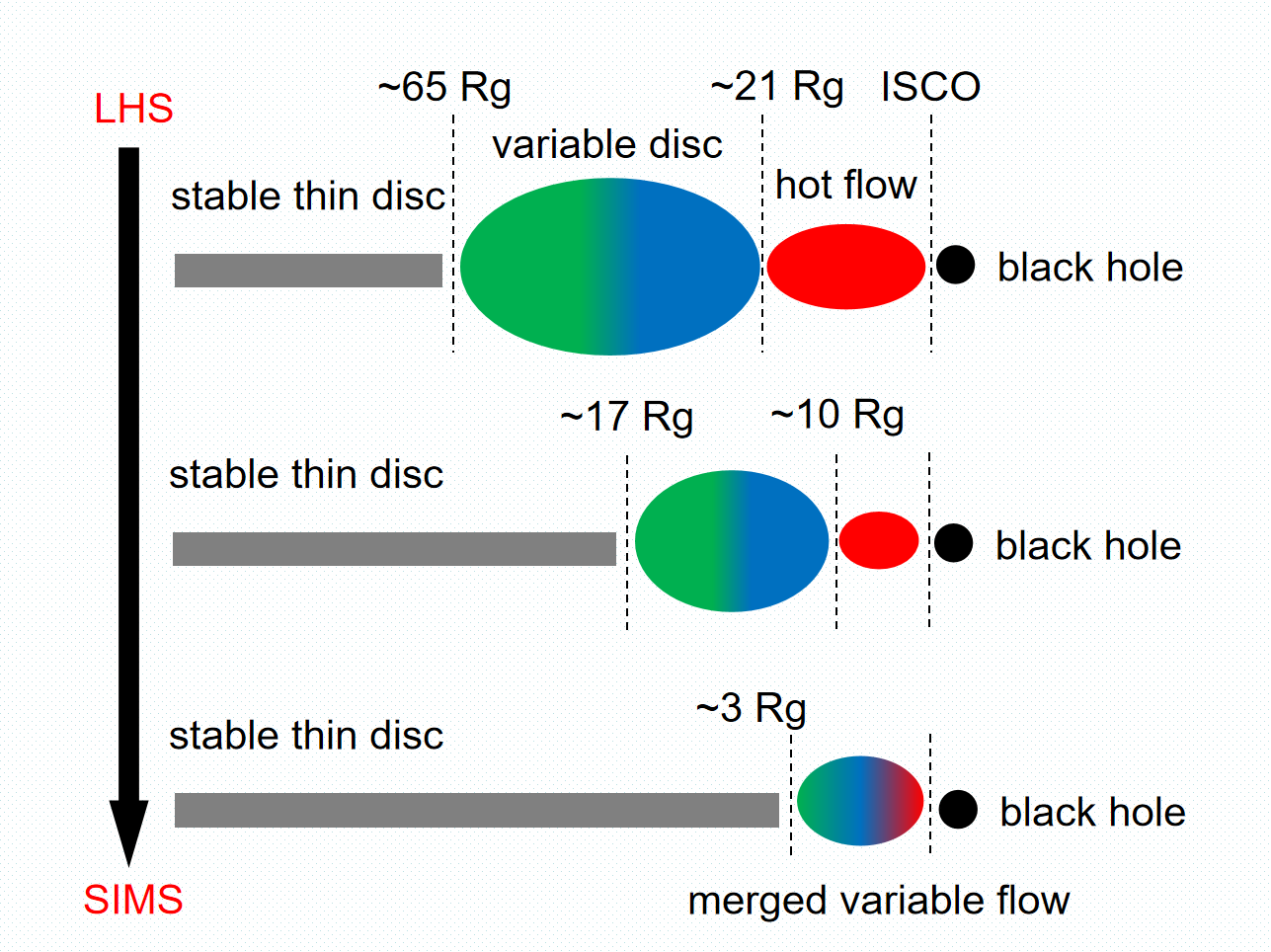}
    \caption{The schematic for the assumed geometry of accretion structure for GX 339-4 from the LHS to the SIMS. The stable thin disc (grey) located far from the black hole is not variable. The fluctuations propagate inwards through the variable flow so the variability with different characteristic frequency is produced in outer region ($L_1$,green) and inner region ($L_2$,blue) of the variable disc and the hot flow ($L_3$,red). When the variable disc closer to the ISCO, the hot flow merges with the inner region of the variable disc into the new merged variable flow.
    }
    \label{fig:geometry}
\end{figure*}

\subsection{Outburst stage}
\label{sec:outburst} 

We made the day-averaged count rate curves of GX 339-4 for its 2021 outburst with \textit{Insight}-HXMT LE data in the Fig.~\ref{fig:light curve}. The count rate showed a consistent trend with the previous work \citep{Jin2023} before MJD 59301, and then stayed stable after MJD 59301. The evolution of hardness as a function of time has also been shown in the Fig.~\ref{fig:light curve}. A rapid decreasing of hardness happened on MJD 59297-59301 and the corresponding rapid change can be also seen in the hardness-intensity diagram (HID, Fig.~\ref{fig:HID}). Hardness is defined as the count rate ratio of 4-8 keV over 2-4 keV, and intensity is defined as the count rate of LE from 2-8 keV. The outburst started at the lower right corner with hardness and intensity changing slightly. After MJD 59297, the outburst rapidly moved to the upper left, and then stayed on upper left after MJD 59301. The rapid decreasing of hardness happened in MJD 59297 indicates the beginning of the state transition from the LHS to the HIMS \citep{Jin2023}. And then the hardness staying stable after MJD 59301 may indicate the outburst entered a new stage (the relatively complete 2021 outburst evolution of GX 339-4 with Insight-HXMT data shown in \citealt{Liu2023}).

For each day, we obtained the PDSs and then fit them with a multiple-Lorentzian model as mentioned in Sect.~\ref{sec:ob}. In Fig.~\ref{fig:qpo_com}, we show the representative PDS fitted with four components. One of the Lorentzian components corresponds to the QPO while three Lorentzian components $L_1$, $L_2$ and $L_3$ are used to represent the broad-band noise components with different characteristic frequencies. The representative PDSs for the different outburst stages are shown in the Fig.~\ref{fig:qpo_stage}. As seen in the left panel, the broad-band noise components dominated the PDSs with the weak QPO in the LHS. In the HIMS, the stronger QPOs and the broad-band noise components with the higher characteristic frequency are shown (the middle panel). However, the broad-band noise and the low-frequency red noise were both found in the PDS of MJD 59301. By studying the PDS with each GTI of MJD 59301, we found that only the broad-band noise and only the red noise were present in the PDS before and after one of the GTIs, respectively. After MJD 59301, the low-frequency red noise with a prominent Type-B QPO was found in the PDSs and the broad-band noise disappeared  (see the right panel of Fig.~\ref{fig:qpo_stage}), which is considered as one of the identifying features of the SIMS \citep{Belloni2016} . Due to the hardness also changing in the same day, we believe the outburst entered the SIMS after MJD 59301. Combined with the state transition from the LHS to the HIMS on MJD 59297, the outburst can be divided to three stage: the LHS (before MJD 59297), the HIMS (MJD 59297 - 59301) and the SIMS (after MJD 59301).
 
\subsection{Evolution of the fractional rms and characteristic frequency with time}
\label{sec:time}

From all the PDSs, we obtained the properties of $L_1$, $L_2$, $L_3$ and QPO components to study their evolution with time (Fig.~\ref{fig:time}). In the top row of panels of Fig.~\ref{fig:time}, we show the evolution of frequency of four components with \textit{Insight}-HXMT LE (left), ME (middle) and HE (right) data. In different energy bands, the QPO frequencies all slightly increase from around 0.1 Hz to 1 Hz during the LHS and then rapidly increase to 5 Hz during the HIMS. Finally, it stays stable around 5 Hz during the SIMS. The characteristic frequencies of $L_1$ and $L_2$ are slightly smaller and larger than the QPOs frequencies, respectively, and both show the similar increasing trend with the QPO frequency until the broad-band noise disappears in the SIMS. However, we found that $L_3$ disappeared before the HIMS, far earlier than $L_1$ and $L_2$, with its characteristic frequency remaining more or less constant around 6 Hz. 

The characteristic frequencies of $L_1$ and $L_2$ increasing with time are consistent with the evolution trend commonly found in other BHXBs \citep{Psaltis1999,Yang2022} . In MAXI J1820-070, all four Lorentzian components used to fit the broad-band noise show the increasing trend with hardness decreasing \citep{Yang2022}. However, for GX 339-4, the characteristic frequencies of the high-frequency component ($L_3$) show no obvious evolution trend in our results. Because the fluctuations are incoherent on time scales shorter than the propagation time \citep{Ingram2011}, the upper limit of the coherent fluctuation is naturally set when the inner radius of the hot flow was very close to the BH. In the bottom three rows of Fig.~\ref{fig:time}, we show the evolution for the fractional rms of $L_1$, $L_2$ and $L_3$. The fractional rms of $L_1$ and $L_2$ shows no obvious trend with time. However, the fractional rms of $L_3$ shows a clear decreasing trend with time in all three energy bands.

In Fig.~\ref{fig:relation}, we study the correlation of the frequency (top row) and fractional rms (bottom rows) between the QPO components and $L_1$, $L_2$ and $L_3$ with \textit{Insight}-HXMT LE data. We fitted the correlation with linear regression in the logarithmic axis to obtain the slope of linear regression curve and Pearson correlation coefficients (PCCs). The PCC of the liner correlation between the characteristic frequency of $L_3$ and QPO frequency is near zero, while those of $L_1$ , $L_2$ are 0.739 and 0.945, respectively. This means that the characteristic frequencies of $L_1$ and $L_2$ show positive linear correlations with QPO frequencies, but the $L_3$ frequency has no correlation with QPO frequencies. However, the fractional rms of three components shows no obvious relationship with the QPOs rms. \cite{Wijnands1999} proposed that a good correlation between the low-frequency break of the broad-band component and the QPO frequency is often observed in BH systems. 

\subsection{Energy dependence of the broad-band noise}
\label{sec:energy}

To study the energy dependence of the different BBN components, we show the PDSs of eight different energy bands in MJD 59380 without the prominent QPO features, as mentioned in Sect.~\ref{sec:ob}. In Fig.~\ref{fig:qpo_energy}, the PDSs of 0.5-8 keV and 10-100 keV are produced with NICER and the \textit{Insight}-HXMT data, respectively. From the top to the bottom,  the PDSs obtained from low energies to high energies are shown. As the energy band increasing, it seems that the low-frequency component has lower fractional rms but the high-frequency component has higher fractional rms. Figure~\ref{fig:energy} shows the energy dependence of the characteristic frequency (top row) and fractional rms (middle row) of $L_1$, $L_2$ and $L_3$. The characteristic frequency of $L_1$ (left) stays stable around 0.1-0.15 Hz while its fractional rms decreases from 30\% to 10\% with energy increasing. In the middle panel, the characteristic frequency of $L_2$ starts increasing from 0.3 Hz below 1 keV, reaches its maximum value of 0.5 Hz around 2 keV and then decreases to 0.4 Hz in the higher energy bands. The fractional rms of $L_2$ shows a similar trend of increasing and decreasing with the maximum value around 2 keV. And this trend also exist in the fractional rms of $L_3$ but with the max value around 6 keV. The characteristic frequency of $L_3$ keeps increasing from 2 Hz to 6 Hz below 6 keV and stays around 6 Hz above 6 keV. 

A similar increasing trend in the characteristic frequency of broad-band components with energy was also found in MAXI J1820+070 \citep{Kawamura2022,Yang2022,gao2024}. \cite{Kawamura2022} found the peak frequency of the high-frequency variability components increase with energy. \cite{Yang2022} found that the characteristic frequency increases with energy for all the broad-band noise components. And a opposite trend of characteristic frequency with energy for low-frequency component and high-frequency component were found by \cite{gao2024}. Although the trend of characteristic frequency for the low-frequency broad-band component was found different, the increasing trend of characteristic frequency for the high-frequency broad-band component was always observed. 
For the energy dependence of the $L_2$ and $L_3$ fractional rms, our result is consistent with the previous work for GX 339-4 \citep{Hitesh2024}. In Fig.5 of \cite{Hitesh2024}, the similar trend of increasing and decreasing with the turning point around 2 and 6 keV for the BF1 (corresponds to $L_2$) and BF2 (corresponds to $L_3$) is shown.

In the bottom row of Fig.~\ref{fig:energy}, we show the energy dependence of the fractional rms ratio of $L_1$, $L_2$ and $L_3$ to study the relation between the broad-band components. In the right panel, the ratio between $L_2$ and $L_3$ shows an increasing trend from 0.6 to 1.15 above 1 keV. And the fractional rms of $L_3$ exceeds that of $L_2$ above 10 keV, which means that $L_3$ is more dominant than $L_2$ in the PDSs. The special point below 1 keV can be explained with the small and similar value of fractional rms for both $L_2$ and $L_3$, which is smaller than that of $L_1$, suggesting that $L_1$ dominates the PDSs in this energy band. The energy dependence of the fractional rms ratio between $L_1$ and $L_2$ (left panel) is more complex. The ratio between $L_1$ and $L_2$ is extremely low ($\sim 0.6$) below 1 keV. With the energy increasing to 1 keV, the ratio comes higher ($\sim 1.5$) but then slightly decreases ($\sim 1.5$) above 10 keV. From 1 to 10 keV, the $L_2$ component is dominant in the PDSs with its the fractional rms larger than $L_1$ and $L_3$. Above 10 keV, the fractional rms of $L_3$ exceeds that of $L_2$ making the PDSs dominated by $L_3$.

\section{Discussion}
\label{sec:dis}
The evolution of the characteristic frequency increasing with time can be well explained with the truncated disc and inner hot flow geometry \citep{Esin1997,Done2007,Ingram2011}. It assumes that the accretion structure consists of a hot, geometrically thick, optically thin inner accretion flow and a cool, geometrically thin, optically thick, outer accretion disc truncated at a larger radius. The outer radius of the hot inner flow is set by the inner radius of the truncated disc. The broad-band noise with a low-frequency and a high-frequency break is believed to originate from the propagating mass accretion rate fluctuations in the hot flow \citep{Lyubarskii1997,Ingram2011,Mushtukov2019,Kawamura2022}. The low-frequency break corresponds to the viscous time-scale at the outer radius of the hot inner flow, but the high-frequency break is more complex than simply the viscous time-scale at the inner radius of the hot flow \citep{Ingram2011}. The Lense–Thirring precession time-scale of the hot flow, which could produce the QPO, is calculated with the combination of inner and outer radius and the surface density of the hot flow. Multiple evidences support the scenario of the inner radius of the disc moving inwards and the contraction of the hot flow from the LHS to the HIMS \citep{Liu2023,Kawamura2023}. As the outer radius of the hot flow decreasing, both the characteristic frequency of the low-frequency broad-band component and QPO frequency increase. Therefore, the observed relation between these two frequencies could be naturally produced. 

\cite{Kawamura2022} developed a model based on the truncated disc and hot flow model with propagating fluctuations, which proposes a thin truncated disc which is not variable and a variable flow including a variable disc and the inner hot flow. In this model, mass accretion rate fluctuations propagate in the variable flow rather than in the hot flow. A difference of the viscous time-scale between the disc and the hot flow produces the double-humps shape of the broad-band noise found in the PDSs. The dip between the double-humps shape corresponds to the drop in viscosity between the two regions. \cite{Kawamura2022} proposed that the geometric size of different accretion regions can be inferred from the shape of the corresponding hump. The low frequency was well fitted by one single standard Lorentzian function probably due to the small radial extent of the variable disc region. And the hot flow may extend in a larger radial range so the intrinsic fluctuations produce Lorentzians of different frequencies in the high-frequency range. In MAXI J1820+070, the high-frequency hump was usually fitted with an asymmetric Lorentzian function \citep{Kawamura2022} or two standard Lorentzian functions \citep{Yang2022,gao2024}. However, the low-frequency hump corresponds to two broad-band components ($L_1$ and $L_2$), while $L_3$ represents the high-frequency hump in our results. This means the accretion geometry of GX 339-4 may be slightly different with a large size of the variable flow and a smaller radial extend range of the inner hot flow (see a schematic of possible accretion flow structures in Fig.~\ref{fig:geometry}). Due to the dip found between the $L_2$ and $L_3$, we consider that the characteristic frequency of $L_1$ corresponds to the viscous time scale at the outer radius of the variable disc, while the inner region of the variable disc and the inner hot flow contribute to the variability of the $L_2$ and $L_3$, respectively. Meanwhile, QPOs with the frequency between $L_1$ and $L_2$ could be explained with the Lense–Thirring precession of the whole variable flow rather than only the inner hot flow.

The decreasing trend of the fractional rms for the high-frequency component was also observed in MAXI J1820+070 \citep{Yang2022,gao2024}.  \cite{Yang2022} attributed the lower value of the fractional rms to the reduction of corona components with the hardness ratios decreasing. This is reasonable because contraction of corona happened from the LHS to the intermediate state with the inner radius of the disc moving inwards to ISCO. With the fractional rms of $L_3$ decreasing from the LHS to the HIMS,  we consider that a similar contraction of corona happened during the evolution in accretion geometry of GX 339-4. However, the fractional rms of the low-frequency component ($L_1$ and $L_2$) shows no obvious evolution with time, which implies that the contribution of the variable disc to the variability may not vary with the change of the variable disc size. 

In Fig.~\ref{fig:radii}, we show a simple quantitative calculation for outer and inner radii of the variable disc based on the standard $\alpha$ disc model \citep{kato,Yang2022} . According to \cite{kato}, the viscous time scale at a disc radius $r$ could be calculated with the viscosity parameter $\alpha$, black hole mass $M_{BH}$ and scale height $H/r$. Given that the characteristic frequencies of broad-band components correspond to the viscous time scales at certain radius, we calculated the outer and inner radii of the variable disc with the characteristic frequencies of $L_1$, $L_2$ and the following parameters: $M_{BH}=10M_{\sun}$, $\alpha=0.1$, $H/r=0.1$\citep{Yang2022,chand2024}. The inner radius of the hot flow was simply assumed at the innermost stable circular orbits (ISCO) of the black hole, because the calculation of the viscous time scale in hot flows is beyond the standard disc model. At the early stage of the outburst, the outer and inner radii of the variable disc were $65R_g$ and $21R_g$, respectively. With the disc moving inwards from the LHS to the HIMS, the outer and inner radii of the variable disc slowly decreased to $17R_g$ and $10R_g$ until the L3 component disappeared. For the other BH candidate MAXI J1820+070, \cite{Kawamura2023} reported a similar shape of PDSs with only the high-frequency hump in the HIMS. They suggested that the hump shape was caused by the variable disc merging with the outer region of inner hot flow. Given that the accretion geometry of GX 339-4 may be slightly different, we suggest that the inner hot flow may merge with the inner region of the variable disc into one region, which could produce the disappearance of the dip and the replacement of $L_3$ by $L_2$ in the high frequency range. After that, the characteristic frequencies of $L_2$ should be related to, but not simply correspond to, the viscous time scale at the inner edge of the merged variable flow, so the inner radius could no longer be calculated based on the standard $\alpha$ disc model. However, the outer radius of the variable flow kept decreasing to $\sim 3R_g$ until broad-band noises disappeared in the SIMS. Considering that GX 339-4 is believed to have a rapid spinning black hole \citep{Parker2016,chand2024}, this radius is very close to its ISCO ($\sim 1R_g$), which supports the scenario of the truncated radius of accretion disc reaching the ISCO in the SIMS.

The energy dependence of the PDSs could be determined by the contribution of different spectral components in each energy band \citep{Kawamura2022}. For example, the extent range (characteristic frequency) of the high-frequency hump is related with the contribution by the hard Comptonisation of the hot flow in MAXI J1820+070 \citep{Kawamura2023}. The larger high-frequency extent region in higher energy band was the result of the more contribution from the hard Comptonisation in the corresponding energy. The spectral results of GX 339-4 \citep{chand2024} show that the energy spectrum was dominated by the the non-thermal radiation of the harder Comptonization component with higher electron temperature and lower optical depth above 10 keV, whereas the origin of photons below 1 keV is mainly the thermal radiation of the multi-colour blackbody component of the disc. This supports the scenario of $L_1$ corresponding to the outer region of the variable disc and $L_3$ corresponding to the inner hot flow, because both $L_1$ and the thermal radiation of the disc are dominant below 1 keV, while above 10 keV both $L_3$ and the non-thermal radiation of the hot flow are dominant. However, the situation for 1-10 keV is more complex due to the photons within this energy band originating from multiple radiation mechanisms, including the thermal radiation of blackbody component, non-thermal radiation of two Comptonization components and the reflected/reprocessed emission. A large optical depth, which is one of the characteristics of the disc component, and low electron temperature were found in the soft Comptonization component. If the innermost region of the disc (corresponding to $L_2$) could produce this soft Comptonization component, it would naturally explains why 1-10 keV is the energy-dominant region of $L_2$. However, this is just a theoretical assumption based on light depth and electron temperature and no evidence has been found for now. A more likely explanation for $L_2$ dominating the PDSs in 1-10 keV is that the seed photons supplied by the inner edge of the disc are able to transfer the variability to other spectral components \citep{Kotov2001,Veledina2016,Veledina2018,Mastroserio2016}. We believe that the results of energy dependence basically support the assumed geometry for the LHS, that is $L_1$, $L_2$ and $L_3$ corresponding to the outer and inner regions of the variable disc and the inner hot flow, respectively (see Fig.~\ref{fig:geometry}). More detailed research on the energy dependence of the broad-band components is still required for studying the radiation mechanism within 1-10 keV to improve the assumed geometry.

In summary, we describe a possible accretion geometry for GX 339-4, which consists of a stable standard thin disc, a big variable disc and a small hot flow in Fig.~\ref{fig:geometry}. As the outburst evolves from the LHS to HIMS, the variable disc region moves inwards to the BH, leading to increasing of the characteristic frequencies of $L_1$ and $L_2$. The evolution of the QPO frequency and its relation with the characteristic frequencies of low-frequency components could also be naturally produced with the decreasing outer radius of the variable disc. Meanwhile, the constant $L_3$ characteristic frequencies and the decreasing trend of the $L_3$ fractional rms indicate that the hot flow may keep contracting but stay close to the ISCO of the black hole. When the inner region of variable disc moves close enough to the black hole, it may merge with the hot flow into one merged region. After that, the merged variable flow would keep moving inwards until it reaches the ISCO in the SIMS. According to \citep{Yang2023}, a possible follow-up may be the merged variable flow converting into a precessing jet, leading to the appearance of Type-B QPOs and the disappearance of broad-band noises. However, the above description about the possible structure of accretion flow is based on the broad-band noise properties and a simple calculation of variable flow regions. Further modelling and data fitting are required to verify and obtain the complete picture of the entire accretion flow structures.

\section{Conclusions}
\label{sec:co}
In this work, we study the outburst of the GX 339-4 with \textit{Insight}-HXMT from 2021 February to March. The transition of outburst stage is studied with the evolution of the count rate, hardness ratio and the power density spectra. Combined with the result of \cite{Jin2023}, the outburst from February to March can be divided to three stages: the LHS (before MJD 59297), the HIMS (MJD 59297 - 59301) and the SIMS (after MJD 59301). In the PDSs of the LHS and the HIMS, broad-band noises and QPOs are found and fitted with multiple-Lorentzian functions. One of the Lorentzian components corresponds to the QPO, and broad-band noises are well fitted with three Lorentzian components: a low-frequency component $L_1$, a middle-frequency component $L_2$ and a high-frequency component $L_3$. We study the evolution of the properties of $L_1$, $L_2$ and $L_3$ with time and their relation to the properties of QPOs.  We also study the energy dependence of $L_1$, $L_2$ and $L_3$ with the \textit{Insight}-HXMT and NICER data. Comparing the fractional rms of the broad-band components with each other, we found that the energy-dominant band of $L_1$, $L_2$ and $L_3$ is below 1 keV, 1-10 keV and above 10 keV, respectively. We suggest a possible accretion geometry  to describe the evolution and energy dependence of QPOs and BBNs based on the truncated disc/inner hot flow geometry with the propagating fluctuations model. Further research and observations are still required to verify this possible accretion structure in GX 339-4.
\begin{acknowledgements}
We are grateful to the referee for the fruitful comments to improve the manuscript. This work is supported by the National Key Research and Development Program of China (Grants No. 2021YFA0718503 and 2023YFA1607901), the NSFC (12133007). 
\end{acknowledgements}

\bibliographystyle{aa}
\bibliography{aa} 

\begin{thebibliography}{60}
\expandafter\ifx\csname natexlab\endcsname\relax\def\natexlab#1{#1}\fi

\bibitem[{Aneesha {et~al.}(2024)Aneesha, Das, Katoch, \& Nandi}]{Aneesha2024}
Aneesha, U., Das, S., Katoch, T.~B., \& Nandi, A. 2024, \mnras, 532, 4486

\bibitem[{Belloni {et~al.}(2005)Belloni, Homan, Casella, van~der Klis, Nespoli, Lewin, Miller, \& Méndez}]{Belloni2005}
Belloni, T., Homan, J., Casella, P., {et~al.} 2005, \aap, 440, 207

\bibitem[{Belloni \& Motta(2016)}]{Belloni2016}
Belloni, T.~M. \& Motta, S.~E. 2016, Astrophysics and Space, 440, 61

\bibitem[{Casella {et~al.}(2005)Casella, Belloni, \& Stella}]{Casella2005}
Casella, P., Belloni, T., \& Stella, L. 2005, \apj, 629, 403

\bibitem[{Chakrabarti {et~al.}(2008)Chakrabarti, Debnath, Nandi, \& Pal}]{Chakrabarti2008}
Chakrabarti, S.~K., Debnath, D., Nandi, A., \& Pal, P.~S. 2008, \aap, 489, L41

\bibitem[{Chakrabarti {et~al.}(2015)Chakrabarti, Mondal, \& Debnath}]{Chakrabarti2015}
Chakrabarti, S.~K., Mondal, S., \& Debnath, D. 2015, \mnras, 452, 3451

\bibitem[{Chand {et~al.}(2024)Chand, Dewangan, Zdziarski, Bhattacharya, Mithun, \& Vadawale}]{chand2024}
Chand, S., Dewangan, G.~C., Zdziarski, A.~A., {et~al.} 2024, \apj, 972, 20

\bibitem[{Chatterjee {et~al.}(2016)Chatterjee, Debnath, Chakrabarti, Mondal, \& Jana}]{Chatterjee2016}
Chatterjee, D., Debnath, D., Chakrabarti, S.~K., Mondal, S., \& Jana, A. 2016, \apj, 827, 88

\bibitem[{Done {et~al.}(2007)Done, Gierliński, \& Kubota}]{Done2007}
Done, C., Gierliński, M., \& Kubota, A. 2007, \aap, 15, 1

\bibitem[{Dziełak {et~al.}(2021)Dziełak, De~Marco, \& Zdziarski}]{Dziełak2021}
Dziełak, M.~A., De~Marco, B., \& Zdziarski, A.~A. 2021, \mnras, 506, 2020

\bibitem[{Esin {et~al.}(1997)Esin, McClintock, \& Narayan}]{Esin1997}
Esin, A.~A., McClintock, J.~E., \& Narayan, R. 1997, \apj, 489, 865

\bibitem[{Gao {et~al.}(2024)Gao, Yu, \& Yan}]{gao2024}
Gao, C., Yu, W., \& Yan, Z. 2024, ArXiv e-prints:2409.06414

\bibitem[{Gendreau {et~al.}(2016)Gendreau, Arzoumanian, Adkins, Albert, Anders, Aylward, Baker, Balsamo, Bamford, Benegalrao, Berry, Bhalwani, Black, Blaurock, Bronke, Brown, Budinoff, Cantwell, Cazeau, Chen, Clement, Colangelo, Coleman, Coopersmith, Dehaven, Doty, Egan, Enoto, Fan, Ferro, Foster, Galassi, Gallo, Green, Grosh, Ha, Hasouneh, Heefner, Hestnes, Hoge, Jacobs, Jørgensen, Kaiser, Kellogg, Kenyon, Koenecke, Kozon, LaMarr, Lambertson, Larson, Lentine, Lewis, Lilly, Liu, Malonis, Manthripragada, Markwardt, Matonak, Mcginnis, Miller, Mitchell, Mitchell, Mohammed, Monroe, Montt~de Garcia, Mulé, Nagao, Ngo, Norris, Norwood, Novotka, Okajima, Olsen, Onyeachu, Orosco, Peterson, Pevear, Pham, Pollard, Pope, Powers, Powers, Price, Prigozhin, Ramirez, Reid, Remillard, Rogstad, Rosecrans, Rowe, Sager, Sanders, Savadkin, Saylor, Schaeffer, Schweiss, Semper, Serlemitsos, Shackelford, Soong, Struebel, Vezie, Villasenor, Winternitz, Wofford, Wright, Yang, \& Yu}]{Gendreau2016}
Gendreau, K.~C., Arzoumanian, Z., Adkins, P.~W., {et~al.} 2016, Proceedings of the SPIE, 9905, 16

\bibitem[{Hynes {et~al.}(2003)Hynes, Steeghs, Casares, Charles, \& O{\textquotesingle}Brien}]{Hynes2003}
Hynes, R.~I., Steeghs, D., Casares, J., Charles, P.~A., \& O{\textquotesingle}Brien, K. 2003, \apj, 583, L95

\bibitem[{Hynes {et~al.}(2004)Hynes, Steeghs, Casares, Charles, \& O{\textquotesingle}Brien}]{Hynes2004}
Hynes, R.~I., Steeghs, D., Casares, J., Charles, P.~A., \& O{\textquotesingle}Brien, K. 2004, \apj, 609, 317

\bibitem[{Ingram \& Done(2010)}]{Ingram2010}
Ingram, A. \& Done, C. 2010, \mnras, 405, 2447

\bibitem[{Ingram \& Done(2011)}]{Ingram2011}
Ingram, A. \& Done, C. 2011, \mnras, 415, 2323

\bibitem[{Ingram \& Done(2012)}]{Ingram2012}
Ingram, A. \& Done, C. 2012, \mnras, 427, 934

\bibitem[{Ingram {et~al.}(2009)Ingram, Done, \& Fragile}]{Ingram2009}
Ingram, A., Done, C., \& Fragile, P.~C. 2009, \mnras, 397, L101

\bibitem[{Ingram {et~al.}(2016)Ingram, Michiel van~der Klis, Middleton, Done, Altamirano, Heil, Uttley, \& Axelsson}]{Ingram2016}
Ingram, A., Michiel van~der Klis, M., Middleton, M., {et~al.} 2016, \mnras, 461, 1967

\bibitem[{Ingram \& van~der Klis(2015)}]{Ingram2015}
Ingram, A. \& van~der Klis, M. 2015, \mnras, 446, 3516

\bibitem[{Ingram \& Motta(2019)}]{Ingarm2019}
Ingram, A.~R. \& Motta, S.~E. 2019, \nar, 85, 101524

\bibitem[{Jin {et~al.}(2023)Jin, Wang, Chen, Tian, Liu, Zhang, Wu, \& Sai}]{Jin2023}
Jin, Y.~J., Wang, W., Chen, X., {et~al.} 2023, \apj, 953, 33

\bibitem[{Kato {et~al.}(2018)Kato, Fukue, \& Mineshige}]{kato}
Kato, S., Fukue, J., \& Mineshige, S. 2018, Black-Hole Accretion Disks — Towards a New Paradigm — (Kyoto University Press)

\bibitem[{Kawamura {et~al.}(2022)Kawamura, Axelsson, Done, \& Takahashi}]{Kawamura2022}
Kawamura, T., Axelsson, M., Done, C., \& Takahashi, T. 2022, \mnras, 511, 536

\bibitem[{Kawamura {et~al.}(2023)Kawamura, Done, \& Takahashi}]{Kawamura2023}
Kawamura, T., Done, C., \& Takahashi, T. 2023, \mnras, 525, 1280

\bibitem[{Kotov {et~al.}(2001)Kotov, Churazov, \& Gilfanov}]{Kotov2001}
Kotov, O., Churazov, E., \& Gilfanov, M. 2001, \mnras, 327, 799

\bibitem[{Liu {et~al.}(2023)Liu, Bambi, Jiang, Garcia, Ji, Kong, Ren, Zhang, \& Zhang}]{Liu2023}
Liu, H., Bambi, C., Jiang, J., {et~al.} 2023, \apj, 950, 5

\bibitem[{Lyubarskii(1997)}]{Lyubarskii1997}
Lyubarskii, Y.~E. 1997, \mnras, 292, 679

\bibitem[{Marcel \& Neilsen(2021)}]{Marcel2021}
Marcel, G. \& Neilsen, J. 2021, \apj, 906, 106

\bibitem[{Markert {et~al.}(1973)Markert, Canizares, Clark, Lewin, Schnopper, \& Sprott}]{Markert1973}
Markert, T.~H., Canizares, C.~R., Clark, G.~W., {et~al.} 1973, \apj, 184, L67

\bibitem[{Mastichiadis {et~al.}(2022)Mastichiadis, Petropoulou, \& Kylafis}]{Mastichiadis2022}
Mastichiadis, A., Petropoulou, M., \& Kylafis, N.~D. 2022, \aap, 662, 9

\bibitem[{Mastroserio {et~al.}(2016)Mastroserio, Ingram, \& van~der Klis}]{Mastroserio2016}
Mastroserio, G., Ingram, A., \& van~der Klis, M. 2016, \mnras, 475, 4027

\bibitem[{Miyamoto {et~al.}(1991)Miyamoto, Kimura, Kitamoto, Dotani, \& Ebisawa}]{Miyamoto1991}
Miyamoto, S., Kimura, K., Kitamoto, S., Dotani, T., \& Ebisawa, K. 1991, \apj, 383, 784

\bibitem[{Molteni {et~al.}(1996)Molteni, Sponholz, \& Chakrabarti}]{Molteni1996}
Molteni, D., Sponholz, H., \& Chakrabarti, S.~K. 1996, \apj, 457, 805

\bibitem[{Mondal {et~al.}(2023)Mondal, Salgundi, Chatterjee, Jana, Chang, \& Naik}]{Mondal2023}
Mondal, S., Salgundi, A., Chatterjee, D., {et~al.} 2023, \mnras, 526, 4718

\bibitem[{Motta {et~al.}(2011)Motta, Mu\~{n}oz Darias, Casella, Belloni, \& Homan}]{Motta2011}
Motta, S., Mu\~{n}oz Darias, T., Casella, P., Belloni, T., \& Homan, J. 2011, \mnras, 418, 2292

\bibitem[{Motta {et~al.}(2015)Motta, Casella, Henze, Mu\~{n}oz Darias, Sanna, Fender, \& Belloni}]{Motta2015}
Motta, S.~E., Casella, P., Henze, M., {et~al.} 2015, \mnras, 447, 2059

\bibitem[{Mushtukov {et~al.}(2019)Mushtukov, Lipunova, Ingram, Tsygankov, Mönkkönen, \& van~der Klis}]{Mushtukov2019}
Mushtukov, A.~A., Lipunova, G.~V., Ingram, A., {et~al.} 2019, \mnras, 486, 4061

\bibitem[{Nowak {et~al.}(2002)Nowak, Wilms, \& Dove}]{Nowak2002}
Nowak, M.~A., Wilms, J., \& Dove, J.~B. 2002, \mnras, 332, 856

\bibitem[{Parker {et~al.}(2016)Parker, Tomsicka, Kennea, Miller, Harrison, Barret, Boggs, Christensen, Craig, Fabian, Fürst, Grinberg, Hailey, Romano, Stern, Walton, \& Zhang}]{Parker2016}
Parker, M.~L., Tomsicka, J.~A., Kennea, J.~A., {et~al.} 2016, \apjl, 821, L6

\bibitem[{Peirano {et~al.}(2023)Peirano, Méndez, García, \& Belloni}]{Valentina2023}
Peirano, V., Méndez, M., García, F., \& Belloni, T. 2023, MNRAS, 519, 1336–1348

\bibitem[{Psaltis {et~al.}(1999)Psaltis, Belloni, \& van~der Klis}]{Psaltis1999}
Psaltis, D., Belloni, T., \& van~der Klis. 1999, \apj, 520, 262

\bibitem[{Remillard \& McClintock(2006)}]{Remillard2006}
Remillard, R.~A. \& McClintock, J.~E. 2006, \araa, 44, 49

\bibitem[{Remillard {et~al.}(2002)Remillard, Sobczak, Muno, \& McClintock}]{Remillard2002}
Remillard, R.~A., Sobczak, G.~J., Muno, M.~P., \& McClintock, J.~E. 2002, \apj, 564, 962

\bibitem[{Schnittman {et~al.}(2006)Schnittman, Homan, \& Miller}]{Schnittman2006}
Schnittman, J.~D., Homan, J., \& Miller, J.~M. 2006, \apj, 642, 420

\bibitem[{Stiele \& Kong(2023)}]{Stiele2023}
Stiele, H. \& Kong, A. K.~H. 2023, \mnras, 552, 268

\bibitem[{Tagger \& Pellat(1999)}]{Tagger1999}
Tagger, M. \& Pellat, R. 1999, \aap, 349, 1003–1016

\bibitem[{Tanenia {et~al.}(2024)Tanenia, Garg, Misra, \& Sen}]{Hitesh2024}
Tanenia, H., Garg, A., Misra, R., \& Sen, S. 2024, \apj, 975, 190

\bibitem[{Uttley(2004)}]{Uttley2004}
Uttley, P. 2004, \mnras, 347, L61

\bibitem[{Uttley \& McHardy(2001)}]{Uttley2001}
Uttley, P. \& McHardy, I.~M. 2001, \mnras, 323, L26

\bibitem[{Uttley {et~al.}(2005)Uttley, McHardy, \& Vaughan}]{Uttley2005}
Uttley, P., McHardy, I.~M., \& Vaughan, S. 2005, \mnras, 359, 345

\bibitem[{van~der Klis(2004)}]{Klis2004}
van~der Klis, M. 2004, Advances in Space Research, 34, 2646

\bibitem[{Varnière \& Tagger(2002)}]{Varnière2002}
Varnière, P. \& Tagger, M. 2002, \aap, 394, 329

\bibitem[{Veledina(2016)}]{Veledina2016}
Veledina, A. 2016, \apj, 832, 181

\bibitem[{Veledina(2018)}]{Veledina2018}
Veledina, A. 2018, \mnras, 481, 4236

\bibitem[{Wijnands \& van~der Klis(1999)}]{Wijnands1999}
Wijnands, R. \& van~der Klis, M. 1999, \apj, 514, 939

\bibitem[{Yang {et~al.}(2022)Yang, Zhang, Bu, Huang, Liu, Yu, Wang, Tao, Qu, Zhang, Zhang, Ma, Song, Jia, Ge, Liu, Yan, Zhou, Li, Wu, Ren, Ma, Zhang, Xu, Ma, Du, \& Xiao}]{Yang2022}
Yang, Z.~X., Zhang, L., Bu, Q.~C., {et~al.} 2022, \apj, 932, 7

\bibitem[{Yang {et~al.}(2023)Yang, Zhang, Zhang, Méndez, García, Huang, Bu, Liu, Yu, Wang, Tao, Altamirano, Qu, Zhang, Ma, Song, Jia, Ge, Liu, Yan, Li, Ren, Ma, Zhang, Xu, Ma, Du, Fu, Xiao, Li, Jin, Zhao, \& Zhao}]{Yang2023}
Yang, Z.-X., Zhang, L., Zhang, S.~N., {et~al.} 2023, \mnras, 521, 3570

\bibitem[{Zhang {et~al.}(2020)Zhang, Li, Lu, Song, Xu, Liu, Chen, Cao, Bu, Chang, Chen, Chen, Chen, Chen, Chen, Cui, Cui, Deng, Dong, Du, Fu, Gao, Gao, Gao, Ge, Gu, Guan, Gungor, Guo, Han, Hu, Huang, Huo, Jia, Jiang, Jiang, Jin, Jin, Li, Li, Li, Li, Li, Li, Li, Li, Li, Li, Li, Liang, Liao, Liu, Liu, Liu, Liu, Liu, Liu, Lu, Lu, Luo, Ma, Meng, Nang, Nie, Ou, Qu, Sai, Shang, Shen, Sun, Tan, Tao, Tuo, Wang, Wang, Wang, Wang, Wang, Wang, Wang, Wen, Wu, Wu, Wu, Xiao, Xiong, Yan, Yang, Yang, Yang, Yi, Yuan, Zhang, Zhang, Zhang, Zhang, Zhang, Zhang, Zhang, Zhang, Zhang, Zhang, Zhang, Zhang, Zhang, Zhang, Zhang, Zhang, Zhang, Zhang, Zhang, Zhang, Zhao, Zhao, Zheng, Zhou, Zhu, Zhu, \& Zhuang}]{Zhang2020}
Zhang, S.-N., Li, T.-P., Lu, F.-J., {et~al.} 2020, Sci. China Phys. Mech. Astron., 63, 249502

\end{thebibliography}

\end{document}